\documentclass[sigconf]{acmart}

\copyrightyear{2026}
\acmYear{2026}
\setcopyright{cc}
\setcctype{by}
\acmConference[SeQureDB '26]{Workshop on Secure and Private Data Management}{May 31-June 05, 2026}{Bengaluru, India}
\acmBooktitle{Workshop on Secure and Private Data Management (SeQureDB '26), May 31-June 05, 2026, Bengaluru, India}
\acmDOI{10.1145/3807894.3810275}
\acmISBN{979-8-4007-2219-6/2026/05}

\newcommand{\sysname}{ANCHOR}

\usepackage{xurl}
\usepackage{subcaption}
\usepackage{graphicx}

\newenvironment{compactItemize}{\begin{list}{\scalebox{0.7}{$\bullet$}}{
    \setlength{\itemindent}{0mm}  
    \setlength{\labelwidth}{2mm} 
    \setlength{\labelsep}{2mm}   
    \setlength{\leftmargin}{4mm} 
}}{\end{list}}

\usepackage{draftwatermark}

\SetWatermarkText{Pre Print. Accepted at SeQureDB 2026}
\SetWatermarkScale{1}
\SetWatermarkAngle{90}
\SetWatermarkFontSize{1cm}
\SetWatermarkHorCenter{27pt}

\begin{document}

\title{\sysname: A Vision for Secure Persistent Key-Value Stores in Disaggregated Data Centers}

\settopmatter{authorsperrow=4}

\author{Viraj Thakkar}
\affiliation{%
  \institution{Arizona State University}
  \city{Tempe}
  \state{AZ}
  \country{USA}
  }
\email{viraj.dt@asu.edu}

\author{Dongha Kim}
\affiliation{%
  \institution{Arizona State University}
  \city{Tempe}
  \state{AZ}
  \country{USA}
  }
\email{dongha@asu.edu}

\author{Hokeun Kim}
\affiliation{%
  \institution{Arizona State University}
  \city{Tempe}
  \state{AZ}
  \country{USA}
  }
\email{hokeun@asu.edu}

\author{Zhichao Cao}
\affiliation{%
  \institution{Arizona State University}
  \city{Tempe}
  \state{AZ}
  \country{USA}
  }
\email{zhichao.cao@asu.edu}


\begin{abstract}

Persistent key--value stores (PKVS) are increasingly deployed in disaggregated settings that split compute, memory, and storage across separate server pools.
This shift redraws the trust boundary: data that would remain within a single machine is now transported, cached, and rewritten across multiple hosts, expanding exposure to both network attackers and intra-infrastructure adversaries.

This paper presents \emph{\textbf{\sysname}}\, a vision for end-to-end integrity and freshness in disaggregated PKVS.
\sysname proposes a two-part semantics-aware architecture: 
1)~\textbf{Persistence path:} \sysname\ outlines encrypting and authenticating PKVS persistent files and preventing rollback with manifest versioning.
2)~\textbf{Volatile path:} \sysname\ treats caches, indexes, and filters as untrusted hints unless accompanied by verifiable provenance, enforced by a TEE-resident policy.
Finally, we outline key invariants and discuss enclave-friendly batching and asynchronous I/O to amortize verification without undermining disaggregation’s performance and elasticity benefits. 

\end{abstract}



\settopmatter{printfolios=true}
\maketitle

\DraftwatermarkOptions{stamp=false}

\section{Introduction}






Persistent key--value stores (PKVS) provide a high-performance interface for persisting service and user data across a wide range of cloud services and storage engines~\cite{cao2020characterizing, glusterfs, rockset, bing_rocksdb_2021, Thakkar2024can}.
To improve scaling and resource utilization, PKVS are increasingly deployed in disaggregated architectures that split compute, memory, and storage across separate server pools~\cite{lin2020disaggregated, dong2023disaggregating, wang2023disaggregated, ewais2024ddc}. 
In these settings, data that once remained within a single machine is now transported, cached, and rewritten across multiple hosts and services~\cite{yu2024caas-lsm, wang_dlsm, lin_o3_2026, legoindex}.
However, this shift redraws the trust boundary: achieving confidentiality and integrity now requires an end-to-end PKVS design that treats the network and infrastructure tiers as first-class parts of the threat surface.


Disaggregation places the adversary within the PKVS data path.
We primarily consider three classes of attackers:
1) an active network attacker that can observe, replay, and modify traffic between server pools;
2) an untrusted backend storage/memory (device, firmware, or service provider) that has physical access to data; and
3) compromised compute-side software, up to an untrusted OS/hypervisor, that can tamper with cached or persistent data.
\emph{Can we preserve the elasticity and performance benefits that motivate disaggregation while ensuring confidentiality and integrity of PKVS data from the above attackers?}

A naive solution is to layer standard protections, such as TLS in transit, data-at-rest encryption, and running sensitive PKVS logic inside a TEE~\cite{perf_tees_ayaz, tls_paper, encryption_popek}.
But a ``protect everything'' approach clashes with the performance and elasticity benefits that motivate disaggregation.
Repetitive encryption adds work to the read/write fast path, limiting performance~\cite{thakkar_shield_2025}.
Also, TEEs have limited trusted memory, so scaling capacity requires mechanisms to manage PKVS caches larger than the TEE's memory, which can become an elasticity bottleneck~\cite{perf_tees_ayaz, IntelSGX}.

\begin{figure*}
    \centering
    \includegraphics[width=0.87\linewidth]{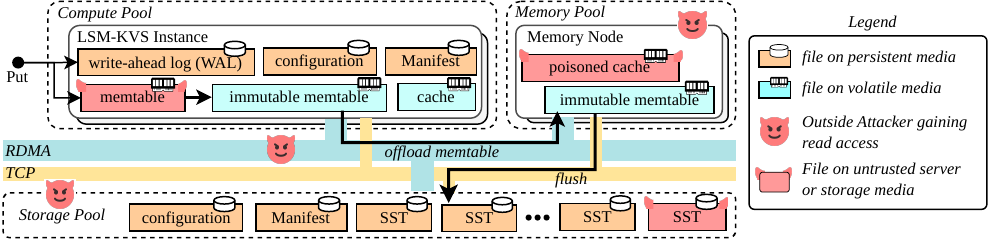}
    \caption{Security Considerations for the Threat Model.}
    \label{fig:threat-model}
\end{figure*}

Given these gaps, recent solutions~\cite{kim2019shieldstore, bailleu2019speicher, shen2023pldb, tweezer} move security into the PKVS rather than treating it as a deployment add-on. 
They encrypt and authenticate PKVS data artifacts (e.g., WAL, SSTables, manifests in certain PKVS~\cite{rocksdb, lsm}) and use a small trusted component (often TEE-backed) to verify integrity and freshness, despite an untrusted host~\cite{bailleu2019speicher, shen2023pldb}. 
However, these designs assume that the PKVS’s in-memory execution context (e.g., caches and filters) is trusted or co-located with a trusted verifier. 
In a disaggregated PKVS, that assumption fails: the memory and compute tiers that provide read/write results can be remote and compromised.

\sysname\ treats disaggregated PKVS as the core premise and assumption, tying security to how PKVS evolves in the new deployment.
To anchor our design, we use log-structured merge-tree-based PKVS (LSM-KVS) as the motivating example for the paper.
LSM-KVS read/write performance depends on continuously evolving in-memory structures (e.g., caches) and persistent data (e.g., WAL, SSTables) that are routinely rebuilt.
Accordingly, our vision is a two-part co-design.
First, persistent objects carry confidentiality and integrity metadata for its contents and evolution step under which they were produced and are valid (e.g., a version descriptor snapshot), enabling detection of rollbacks and mix-and-match across WAL/SST histories even when individual objects remain well-formed.
Second, a small trusted mediator controls admission into caches and derived metadata and checks provenance on use: any value returned to applications must be derivable from authenticated persistence, even if the OS/hypervisor and remote memory tier are compromised.
Using LSM disaggregation as a concrete thread, the rest of the paper sketches mechanisms and invariants that preserve disaggregation’s elasticity without sacrificing confidentiality of stored data and end-to-end integrity.

\smallskip\noindent
To summarize, we envision PKVS security by leveraging:
\begin{itemize}
  \item Integrity (and freshness) must bind to both contents and the PKVS evolution contract to prevent rollback and mix-and-match.
  \item Caches, indexes, and filters are untrusted; returned values must be derivable from authenticated persistence.
  \item Trusted state must be small and exposed restricted interfaces so the trusted mediator does not become an elasticity bottleneck.
  \item Asynchronous I/O and batching can amortize verification costs.
\end{itemize}

\DraftwatermarkOptions{stamp=true}

\section{Background}
\label{sec:background}






PKVS exposes a simple, high-performance interface (e.g., \texttt{Put}, \texttt{Get}, \texttt{Delete}) and serves as the persistent layer for caches, metadata services, and storage engines~\cite{bing_rocksdb_2021, cao2020characterizing, bigtable, rockset, zippydb, leveldb, rocksdb}.
Different PKVS designs employ different underlying data structures, each with its own trade-offs.
For instance, typically, B$+$-tree-based engines demonstrate better read, while log-structured merge-tree-based (LSM) engines demonstrate better write performance~\cite{lsm, btree}.
We use an LSM-based PKVS (LSM-KVS) as the example for this paper.

\smallskip\noindent
\textbf{Data flow in LSM-KVS.}
In a conventional single-node deployment, each write is appended to a write-ahead log (WAL) for durability and inserted into an in-memory memtable.
When the memtable fills, it is flushed into an immutable, sorted on-disk file (an SSTable), and the system records the current set of live SSTables and key ranges in a versioned metadata structure (often called a manifest or version set).
Reads consult the memtable (and any immutable memtables), then use execution-derived in-memory structures (block cache, block indexes, Bloom filters, and per-level metadata) to identify candidate SSTables and fetch blocks from local storage. 
In the background, compaction continuously rewrites SSTables to maintain read performance and reclaim space from overwritten keys and tombstones~\cite{lsm, rocksdb, leveldb}.
Critically, this pipeline was designed under a \emph{trusted-node} assumption: keys/values and metadata are typically handled as plaintext within the machine boundary, and integrity mechanisms (when present) are primarily aimed at accidental corruption (e.g., checksums), not adversarial modification~\cite{checksum_blog, checksum_github}.

\smallskip\noindent
\textbf{LSM-KVS in Disaggregated Data Centers.}
Disaggregation separates the LSM-KVS compute from the resources that hold its state: compute nodes ingest requests and maintain memtables, while WAL segments and SSTables are persisted to a remote storage service (e.g., over a storage fabric or object interface), and caches and read accelerators may be externalized to a shared remote memory tier~\cite{wang_dlsm, lin_o3_2026}.
Flush and compaction still generate new SSTables and update manifests, but now these artifacts are \emph{transported, cached, and re-read across multiple hosts and services}~\cite{yu2024caas-lsm, dong2023disaggregating}, often with retries and replication as part of normal operation.
Without end-to-end confidentiality and integrity, as shown in \figurename~\ref{fig:threat-model}, disaggregation amplifies exposure in two ways: \textbf{1)} data and metadata become visible and mutable \emph{in transit} and at the remote service, enabling replay/rollback and mix-and-match of WAL/SST/manifest histories; and \textbf{2)} the execution-derived state that determines read outcomes (caches, filters, indexes) can be poisoned or desynchronized from persistent state, causing incorrect reads even if the underlying storage bytes are merely ``well-formed.''
These observations motivate security mechanisms that bind correctness to the LSM-KVS and constrain what untrusted tiers may cache, rewrite, or serve.

\section{Motivation}
\subsection{Threat Model}
\label{sec:threat-model}

We aim to maintain the confidentiality and integrity of the PKVS state in disaggregated deployments.
Our threat model reflects the fact that disaggregation places untrusted components \emph{on the critical path} of both foreground operations (reads/writes) and background maintenance (flush/compaction).

\begin{table*}[]
    \centering
    \small
    \captionsetup{skip=2pt}
    \caption{LSM-KVS Components and Security Properties Required.}
    \label{tab:lsm-component-security}
    \begin{tabular}{lllll}
        \toprule
        \multicolumn{1}{c}{\textbf{Data File / State}} &
            \multicolumn{1}{c}{\textbf{Present on (Disagg. Pool)}} &
            \multicolumn{1}{c}{\textbf{Primary Use}} &
            \multicolumn{1}{c}{\textbf{Attack Points}} &
            \multicolumn{1}{c}{\textbf{Needed Property}} \\
        \midrule
        Write-Ahead Log 
            & compute, storage 
            & durability  
            & rollback, truncation  
            & integrity + monotonicity \\
        Memtable
            & compute, memory
            & writes
            & substitution, mix-and-match 
            & admission control + authenticity \\
        SSTable
            & storage 
            & reads 
            & substitution, mix-and-match 
            & authenticity + binding to version \\
        Manifest
            & compute, memory, storage 
            & defines latest file  
            & rollback
            & freshness + append-only history   \\
        Block cache 
            & compute, memory  
            & latency 
            & poisoning, staleness 
            & admission control + provenance    \\
        Bloom filter / Index
            & compute, memory
            & candidate pruning 
            & false negatives / misdirect reads 
            & authenticated derivability \\
        \bottomrule
    \end{tabular}
\end{table*}

\smallskip\noindent\textbf{Assets.}
We treat the following as sensitive:
1) any file holding key--value contents (e.g., WAL and SSTable), and 
2) the metadata that determines visibility and recovery (e.g., manifests).
Because LSM performance depends on caching and derived metadata, we also consider transient replicas on the compute side (block caches, compaction working sets, and derived structures) as part of the attack surface, even if they are not persistent.

\smallskip\noindent\textbf{Adversaries and capabilities.}
We utilize threat models from prior research on LSM-KVS security (e.g., TWEEZER~\cite{tweezer}, SPEICHER~\cite{bailleu2019speicher}) and adapt them to account for the requirements of Disaggregation. We consider three attacker classes.
\begin{compactItemize}
    \item \textbf{Active network attacker:} can observe, drop, delay, reorder, replay, and modify traffic between client/compute/storage pools.
    \item \textbf{Untrusted backend services (remote storage/memory):} can read/modify PKVS state hosted outside the compute node, including persistent artifacts (WAL, SSTables, manifests) and remote derived state (cache entries, Bloom filters); it can return stale snapshots and can attempt to poison ``hint'' structures to bias reads (e.g., induce false negatives).
    \item \textbf{Compromised compute-side software:} the OS/hypervisor and surrounding software on compute nodes may be malicious, allowing attackers to tamper with caches, metadata, and I/O behavior; in particular, attempt rollback or mix-and-match attacks.
\end{compactItemize}

\smallskip\noindent
\textbf{Assumptions and non-goals.}
We assume sound cryptographic primitives/key management and correct execution of the TEE-backed mediator.
We assume a key distribution service such as Kerberos~\cite{neuman1994kerberos} or Secure Swarm Toolkit~\cite{kim2017toolkit, kim2023sst}, for key provisioning across nodes.
We do not address denial-of-service, access-pattern, size, timing leakage, or microarchitectural/physical attacks; we focus on integrity and freshness under adversarial infrastructure.

\smallskip\noindent
\textbf{Trust boundary.} 
The only trusted component is the \sysname\ mediator (TEE-backed) that holds cryptographic keys and validates artifact provenance/version evolution. The OS/hypervisor, the rest of the PKVS process, the remote memory tier, and the storage service may behave adversarially.

\subsection{Motivation}

In a monolithic deployment, a PKVS can often treat the storage device as a relatively stable base and rely on the local kernel and file system to provide basic mediation.
Disaggregation removes these assumptions: persistent objects are fetched from remote services that may be adversarial, staged through caches that may outlive the moment they were fetched, and continuously rewritten by asynchronous maintenance (flush/compaction) decoupled from foreground requests. 
The security question, therefore, shifts from ``are the bytes encrypted'' to ``does every query result reflect an authenticated and current logical state of the PKVS.''

Two properties make this sharp for LSM-KVS. 
\textbf{\textit{First,}} correctness depends on \emph{versioned evolution}: a storage adversary can replay an older manifest or serve stale-but-well-formed WAL/SSTables, inducing rollback or mix-and-match histories without triggering naive integrity checks. 
\textbf{\textit{Second,}} performance hinges on \emph{derived state} (caches, indexes, filters) that steers reads; if poisoned or stale, it can bias execution and make incorrect results sticky, especially when the OS/hypervisor or remote memory tier can tamper with buffer provenance and I/O completion.

Existing secure LSM-KVS designs, such as TWEEZER~\cite{tweezer} and PLDB~\cite{shen2023pldb}, authenticate core LSM artifacts (e.g., SSTables, WAL, manifest) and use a TEE-backed engine to detect rollback-style attacks. 
However, their security boundary largely assumes a single trusted engine rather than disaggregated deployments, in which the data is typically outside those boundaries.
Consequently, even if individual persisted objects verify, an adversary can still influence \emph{which} authenticated objects are used (or suppress required reads) unless the PKVS also binds execution-relevant state to a monotonic version timeline and checks its provenance.

These observations also sketch why mechanisms layered below or above the PKVS are insufficient in isolation. 
At-rest encryption~\cite{encryption_popek, rocksdb-encryption, hadoop_transparent_encryption} and transport security~\cite{tls_paper} protect channels and media but do not prevent rollback/equivocation by a compromised backend, nor do they constrain derived state. 
Generic storage-layer authentication can validate blocks, but without binding validation to PKVS versions and object lifecycles, it cannot answer the question the PKVS must answer on every read: ``is this the correct object for the current logical state?'' 
Conversely, pushing cryptography entirely to applications protects values but often sacrifices server-side indexing and does not ensure the PKVS faithfully enforces freshness and consistency under adversarial persistence.

\section{Approach}

\sysname\ operationalizes LSM-KVS requirements by co-designing cryptographic protection with LSM artifact semantics and by constraining the in-memory interpretation path under an untrusted OS and remote memory tier. The key idea is to make the version descriptor (manifest/version set) the root of authenticity for the engine’s logical state, and to ensure that (i) every persisted object is bound to that evolving logical state, and (ii) every cached byte that can influence query outcomes carries verifiable provenance.

\begin{figure*}[]
  \centering
  \begin{subfigure}[]{0.48\linewidth}
    \centering
    \includegraphics[width=0.8\textwidth]{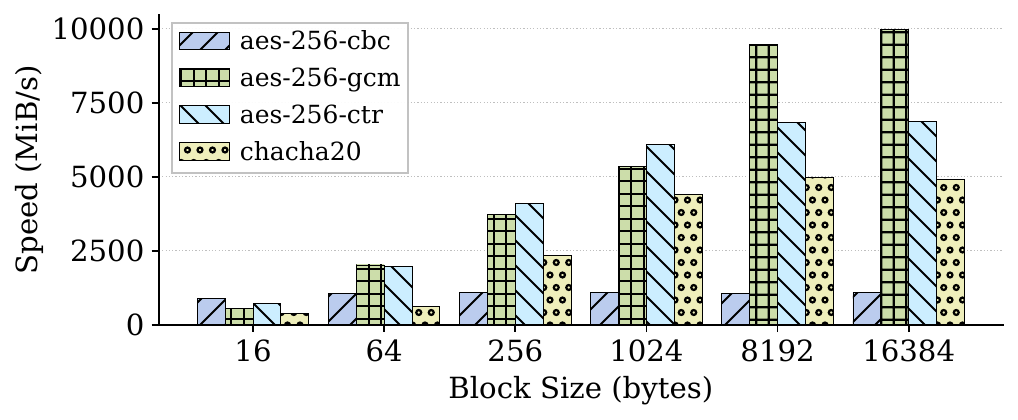}
    \vspace{-2pt}
    \caption{Encryption microbenchmark. Throughput (MiB/s) across block sizes for AES-{CBC,GCM,CTR} and ChaCha20; used to reason about block-granular authentication vs record-granular WAL protection}
    \label{fig:2a}
  \end{subfigure}
  \hfill
  \begin{subfigure}[]{0.46\linewidth}
    \centering
    \includegraphics[width=0.8\textwidth]{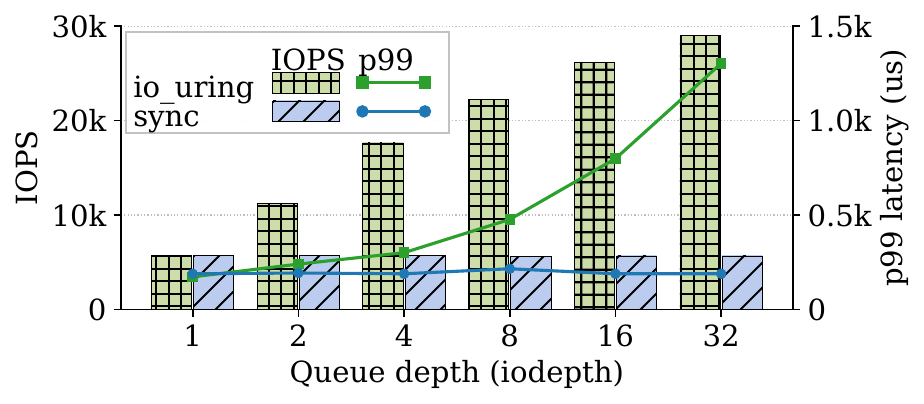}
    \vspace{-2pt}
    \caption{Async I/O microbenchmark. io\_uring vs sync submission; reports IOPS and p99latency vs queue depth; used to motivate batching verification and reducing enclave transitions.}
    \label{fig:2b}
  \end{subfigure}
  \caption{Co-design knobs and preliminary evidence s for verify-before-use in disaggregated LSM-KVS.}
  \label{fig:codesign-eval}
\end{figure*}

\subsection{Version-bound Authenticated Persistence}

\sysname\ starts from the observation (Table ~\ref{tab:lsm-component-security}) that LSM-KVS data differs in lifecycle (append vs rewrite), granularity (record vs block vs file), and security requirements. 
These differences justify different cryptographic constructions and verification granularities—but not different security goals. 
In our setting, every artifact that can change read outcomes must be integrity-protected against an active adversary; confidentiality is applied where it protects meaningful secrets (e.g., values and, optionally, sensitive metadata).

\sysname\ centers the version descriptor (manifest/version set) as the state root: it commits to which objects are live and to the cryptographic commitments needed to validate them. This turns Table 1’s “binding to version” into a concrete organizing principle: an SST block is not “valid” because its bytes authenticate; it is valid because it authenticates as part of the currently installed version.

This framing leads to an artifact-aware protection policy:
\begin{itemize}
    \item WAL segments (append-only stream). Use a streaming-friendly authenticated format that preserves integrity $+$ monotonicity for record sequences. This favors constructions that batch well and amortize per-record overhead.
    \item SSTables (random reads, compaction rewrites). Protect at block granularity with authentication that binds blocks to their owning file and the referencing version descriptor, preventing substitution or cross-version splicing. 
    \item Manifest/version descriptors (state-defining). Protect freshness with an append-only evolution rule (e.g., ``latest'' anchor in the trusted mediator), making rollback and fork detectable as violations of the PKVS timeline rather than as generic corruption.
\end{itemize}
Why granularity is a systems knob. The authentication unit (record/ block/file) directly changes CPU cost per I/O and therefore how aggressively ANCHOR can verify before use. Figure~\ref{fig:2a} illustrates this effect: encryption throughput varies sharply with block size and algorithm, implying that ‘secure-by-default’ choices that are fine for large SST blocks can be disproportionately expensive for small WAL records. This motivates a semantics-driven selection: streaming-friendly authentication for WAL, and block-granular authentication for SSTables where reads already occur in blocks.”

\subsection{Constrained In-Memory Execution}

Even if persistent objects are protected, disaggregation still introduces a correctness hazard: derived state can silently steer execution. Block caches, filters, and indexes decide which objects are fetched and merged; poisoning or staleness can make incorrect data “sticky,” especially when the OS and remote memory tier are adversarial.

This shifts caches/accelerators from being implicit trust anchors into being untrusted hints. Two practical consequences follow:
\begin{compactItemize}
    \item Cache admission $+$ provenance. A cache hit is not trusted because it may be poisoned; entries must carry a provenance label (e.g., version descriptor authorization). The mediator can reject stale or foreign entries, converting poisoning into safe misses.
    \item Filters/indexes must either be authenticated or non-authoritative. If a Bloom filter or index is authenticated, it can be used for pruning. Otherwise, it is treated as a hint that may add work but cannot exclude candidates, preventing silent correctness violations from becoming false negatives.
\end{compactItemize}

Finally, \sysname\ must reconcile this verification discipline with performance. Disaggregation’s throughput benefits depend on asynchronous I/O and batching, yet TEEs tend to penalize syscall-heavy paths and frequent enclave transitions. \sysname’s vision is to separate untrusted I/O submission from trusted admission: the OS (or helper threads) can drive high-throughput async I/O (e.g., via io\_uring~\cite{didona2022understanding}), but the mediator is the sole authority that turns completed I/O into usable PKVS state. The central systems question is whether we can obtain io\_uring-like efficiency while keeping the trusted mediator small and minimizing enclave transitions through batching and pipelining.

Figure~\ref{fig:2b} provides preliminary evidence that asynchronous submission can increase IOPS while controlling tail latency as queue depth grows, which is exactly the operating regime where mediator-side verification must batch work to avoid per-request enclave transitions. 
In ANCHOR, this suggests a split-phase design: untrusted threads drive deep async I/O, while the mediator verifies completions in batches and admits only verified blocks.

\vspace{-2pt}
\section{Discussion}

The vision surfaces several co-design questions at the boundary of storage semantics, cryptography, and disaggregated systems:

\begin{compactItemize}
    \item Cost models for artifact-aware protection: how should the PKVS choose between constructions (e.g., per-record vs per-chunk authentication, per-block vs per-file commitments) under workload skew and compaction intensity?
    \item Trusted admission at scale: what is the minimal provenance metadata that prevents stale/poisoned derived state without turning the mediator into an elasticity bottleneck?
    \item Enclave-friendly async verification: how can batching and split-phase I/O reduce enclave transitions while preserving verify-before-use under an untrusted OS?
    \item Compaction under partial trust: can compaction be offloaded outside the trusted boundary while still making the resulting files verifiably consistent with the version evolution contract?
\end{compactItemize}





\section{Conclusion}

ANCHOR's vision is to restore end-to-end confidentiality and integrity by co-designing security with PKVS semantics: 
1)~bind WAL/SST/manifest integrity to a version-evolution timeline (with manifest as authenticity root), and 
2)~constrain in-memory interpretation path so caches and derived metadata are treated as untrusted hints until provenance is verified. 
This separation enables a small trusted mediator to enforce verify-before-use while leveraging asynchronous I/O to preserve disaggregation's performance benefits.

We take the first step towards this vision in SHIELD~\cite{thakkar_shield_2025}.
Key next steps are to quantify the cost trade-offs of artifact-aware protection, minimize provenance state for scalable admission control, and support enclave-friendly verification and compaction offload without breaking the version-evolution contract.

\begin{acks}
We would like to thank our anonymous reviewers for their valuable feedback. We thank all members of the ASU-IDI Lab for their help and useful comments. This work was partially funded by the National Science Foundation (NSF) under Grant Numbers \href{https://www.nsf.gov/awardsearch/showAward?AWD_ID=2412436}{\#2412436}, \href{https://www.nsf.gov/awardsearch/showAward?AWD_ID=2443219}{\#2443219}, and \href{https://www.nsf.gov/awardsearch/show-award/?AWD_ID=2449200}{POSE-\#2449200}.
\end{acks}

\bibliographystyle{ACM-Reference-Format}
\bibliography{shield_refs, zhichao-references}

\end{document}